\documentclass[fleqn, usenatbib]{mnras}

\usepackage{newtxtext,newtxmath}
\usepackage[T1]{fontenc}
\DeclareRobustCommand{\vAN}[3]{#2}
\let\vANthebibliography\thebibliography
\def\thebibliography{\DeclareRobustCommand{\vAN}[3]{##3}\vANthebibliography}

\usepackage{graphicx}	
\usepackage{subfig}
\usepackage{amsmath}	
\usepackage{float}
\usepackage{physics}
\usepackage[latin1]{inputenc}
\usepackage[dvipsnames]{xcolor}
\usepackage{multicol}

\newcommand{\msun}{M_{\odot}}
\newcommand{\ff}{_{\mathrm{ff}}}
\newcommand{\eff}{\epsilon\ff}
\newcommand{\crit}{\mathrm{cr}}
\newcommand{\G}{\mathrm{G}}
\newcommand{\pc}{\mathrm{pc}}
\newcommand{\cm}{\mathrm{cm}}
\newcommand{\gas}{_{\mathrm{gas}}}
\newcommand{\SFR}{_{\mathrm{SFR}}}
\newcommand{\lf}{\ell_{\mathrm{f}}}
\newcommand{\vA}{v_{\mathrm{A}}}
\newcommand{\Ma}{\mathcal{M}_{\mathrm{A}}}
\newcommand{\Mf}{\mathcal{M}_{\mathrm{f}}}
\newcommand{\Ms}{\mathcal{M}_{\mathrm{s}}}
\newcommand{\cs}{c_{\mathrm{s}}}
\newcommand{\cf}{c_{\mathrm{f}}}

\definecolor{myblue}{RGB}{30, 136, 229}

\title[A new magneto-hydrodynamic star formation recipe]{A new star formation recipe for magneto-hydrodynamics simulations of galaxy formation}

\author[E. Girma et al.]{
Eden Girma,$^{1}$\thanks{E-mail: egirma@princeton.edu}
Romain Teyssier,$^{1}$
\\
$^{1}$Department of Astrophysical Sciences, Princeton University, Peyton Hall,
Princeton, NJ 08544, USA
}

\date{Accepted XXX. Received YYY; in original form ZZZ}

\pubyear{2022}

\begin{document}
\label{firstpage}
\pagerange{\pageref{firstpage}--\pageref{lastpage}}
\maketitle

\begin{abstract}
Star formation has been observed to occur at globally low yet locally varying efficiencies.  As such, accurate capture of star formation in numerical simulations requires mechanisms that can replicate both its smaller-scale variations and larger-scale properties. Magnetic fields are thought to play an essential role within the turbulent interstellar medium (ISM) and affect molecular cloud collapse.  However, it remains to be fully explored how a magnetised model of star formation might influence galaxy evolution.  We present a new model for a sub-grid star formation recipe that depends on the magnetic field. We run isolated disk galaxy simulations to assess its impact on the regulation of star formation using the code \texttt{RAMSES}.  Building upon existing numerical methods, our model derives the star formation efficiency from local properties of the sub-grid magnetised ISM turbulence, assuming a constant Alfv\'en speed at sub-parsec scales.  Compared to its non-magnetised counterpart, our star formation model suppresses the initial starburst by a factor of two, while regulating star formation later on to a nearly constant rate of $\sim 1~\msun~\mathrm{yr}^{-1}$. Differences also arise in the local Schmidt law with a shallower power law index for the magnetised star formation model.  Our results encourage further examination into the notion that magnetic fields are likely to play a non-trivial role in our understanding of star and galaxy formation. 
\end{abstract}




\section{Introduction}
Star formation in disk galaxies results from a complex interaction between various mechanisms, such as gravitational collapse, cooling, and turbulence. However, a comprehensive model of both the sum and its parts has yet to be wholly established.  Star-forming regions in spiral, gas-rich galaxies generally convert available gas to stars at a surprisingly low instantaneous efficiency \citep{persic_baryon_1992, fukugita_cosmic_1998, balogh_revisiting_2001}, with numerical simulations and observational constraints suggesting a parent dark matter halo mass dependence that peaks at $\sim$20\% efficiency for Milky Way sized objects \citep{behroozi_lack_2013,behroozi_average_2013}.  These regions can also be characterised by the ratio of available gas mass to star formation rate, otherwise known as the depletion time scale $\tau_{\mathrm{d}}$ \citep{leroy_star_2008, bauermeister_gas_2010}.  

While observed star formation rates can vary dramatically across disk galaxies, empirical correlations emerge at scales larger than $100$~pc. The \citet{schmidt_rate_1959} law describes the relation between observed star formation rate surface density $\Sigma_{\mathrm{SFR}}$ and gas surface density $\Sigma_{\mathrm{gas}}$ as a simple power law,
\begin{equation}
    \Sigma_{\mathrm{SFR}} = A\Sigma^{n}_{\mathrm{gas}}
\end{equation}
where $A$ is an empirically determined normalisation constant. Local Schmidt laws, as measured radially within individual galaxies, demonstrate power law correlations with widely ranging slopes and normalisations \citep{wong_relationship_2002}. However \citet{kennicutt_star_1998} first found that when averaged over the entire galaxy, a global Kennicutt-Schmidt relation arises and appears consistent across galaxies, following a power law with slope $n \approx 1.4$. 

It remains unclear what physical processes, or interplay of processes, result in these observed relations, though presently debated models invoke the role of gravitational instabilities \citep[e.g.][]{quirk_gas_1972, larson_large-scale_1988, elmegreen_supercloud_1994, kennicutt_star_1989, kennicutt_star_1998}, cloud-cloud collisions \citep[e.g.][]{wyse_schmidt_1986, wyse_star_1989, silk_feedback_1997, tan_star_2000}, turbulence and its influence on the gas density probability distribution function (PDF) \citep[e.g.][]{scalo_probability_1998, passot_density_1998, ostriker_kinetic_1999, klessen_one-point_2000, wada_numerical_2001, ballesteros-paredes_physical_2002, li_formation_2003, kravtsov_origin_2003, mac_low_distribution_2005}, and gas thermodynamics \citep[e.g.][]{struck-marcell_hydrodynamic_1991, struck_simple_1999}. Due to the difficulty of encoding all these physics into cosmological simulations, as well as the inability to resolve individual molecular clouds or multiphase ISM, galaxy formations frequently assume a sub-grid star formation recipe based on the Schmidt law.  In absence of an explicit Schmidt-like star formation prescription, however, Kennicutt-Schmidt relations have successfully emerged in more recent stratified disk models \citep{li_star_2006, bieri_satin_2023}.  These models notably suggest that a dynamic interplay between gravitational instabilities, supernovae (SNe) feedback, and gas cooling somehow drives star formation rates towards what is locally observed.

Schmidt-like sub-grid star formation recipes can assume a uniform and constant $\eff$, the value of which can be adjusted to reproduce the observed Kennicutt-Schmidt relation \citep[see e.g.][]{li_star_2006, wada_density_2007}. The star formation rate can then be modelled as
\begin{equation}
    \dot{\rho}_{*} = \eff \frac{\rho\gas}{t\ff}; \; t\ff = \sqrt{\frac{3\pi}{32G\rho\gas}}
\end{equation}
where $\rho\gas$ is the local gas density and $t\ff$ is the local free-fall time \citep{stinson_star_2006}. However, extreme variation of $\eff$ has been observed in individual molecular clouds of radii $R_{\mathrm{MC}}\lesssim 100$ pc, with values ranging from $1$\% to $30$\% \citep{murray_star_2011}.  A physical model for star formation must necessarily explain and capture this smaller-scale variability, while accurately reproducing the observed larger-scale relations and the overall low efficiency in galaxies.

Attempts to identify what external and internal factors regulate star formation have long been underway.  Several authors have explored supersonic turbulence in the interstellar medium (ISM) as a promising candidate, as it both counteracts gravitational collapse via turbulent pressure and induces gravitational collapse by sporadically creating dense regions \citep{chandrasekhar_fluctuations_1951, bonazzola_jeans_1987, krumholz_general_2005, hennebelle_analytical_2008, federrath_star_2012}.  The presence of turbulence in molecular clouds can be inferred from the empirical proportionality between velocity dispersion and cloud size \citep{larson_turbulence_1981} as well as from measurements of velocity and density power spectra \citep{heyer_universality_2004}.  

Yet without any driving forces the turbulent energy decays far too quickly relative to the longest estimates of molecular cloud lifetimes \citep{blitz_origin_1980,stone_dissipation_1998, mac_low_kinetic_1998}.  Potential mechanisms for turbulence production in molecular clouds exist on both large scales ($\gtrsim R_{\mathrm{MC}}$) and small scales ($\lesssim R_{\mathrm{MC}}$). Models in which larger scale processes like SN feedback and spiral shock forcing dominate turbulence driving appear most consistent with observations \citep{brunt_turbulent_2009}. Turbulence in the ISM may also be driven at smaller scales by stellar outflows \citep{bally_parsec-scale_1996, knee_molecular_2000}, stellar winds \citep{norman_clumpy_1980, lada_energetic_1982, franco_self-regulated_1983}, and compact photoionizing HII regions \citep{matzner_role_2002}.

One factor that remains significant, and is known to operate on a variety of scales as both a regulator and a trigger of star formation is the magnetic field. Magnetic fields are capable of supporting molecular clouds up to a critical mass value \[M_\crit = 0.13 \frac{\Phi}{G^{1/2}} \approx 10^3~\msun \left(\frac{B}{30~\mu\G} \right) \left( \frac{R}{2~\pc} \right)^2\] where $\Phi$ is the magnetic flux, $B$ is the magnetic field strength, and $R$ is the molecular cloud radius \citep{mouschovias_note_1976, shu_star_1987}. For clouds of mass $M< M_\crit$ (known as "subcritical"), ambipolar diffusion can trigger fragmentation and star formation within their interior by adequately redistributing magnetic flux \citep{mestel_star_1956,mouschovias_nonhomologous_1976,mouschovias_nonhomologous_1976-1,mouschovias_star_1987}. As clouds with sufficiently low ion-neutral collisions settle into a gravitationally unstable neutral core and ionized envelope, magnetic fields regulate timescales of collapse by virtue of the ambipolar diffusion timescale, \[\tau_{\mathrm{AD}} \lesssim 7.3~\mathrm{Myr}~\left(\frac{n_i}{10^{-1}~\cm^{-3}}\right) \left(\frac{n_c}{10^5~\cm^{-3}} \right)^{-2/3} \left( \frac{M_c}{10^2~\msun}\right)^{2/3}\] where $n_i$ is the ion density, $n_c$ is the neutral core density, and $M_c$ is the neutral core mass \citep{mouschovias_connection_1977}.

However, this process requires strong initial magnetic fields to explain rapid star formation rates observed over $\lesssim 1~$Myr \citep{hartmann_rapid_2001}. In addition, most recent observations suggest that the median molecular clouds are actually "supercritical", where magnetic fields alone cannot counteract collapse \citep{crutcher_magnetic_2012}. For these clouds, magnetised supersonic turbulence appears to decrease star formation by factors of 2-3 when compared with non-magnetised turbulent flows
\citep{price_inefficient_2009, dib_angular_2010, padoan_star_2011, federrath_star_2012, padoan_simple_2012}. Reasons for this seem to be at least twofold.  Simulations have demonstrated that the presence of magnetic fields narrows the PDF imposed on gas density by turbulent flows, making it more difficult to reach higher densities \citep{cho_compressible_2003, kowal_density_2007, burkhart_density_2009, molina_density_2012, mocz_moving-mesh_2017}.  Furthermore, magnetic fields change the criteria for collapse itself, as they provide additional support against self-gravity. 

Past studies have added magneto-hydrodynamics (MHD) to turbulence-regulated star formation theoretical models \citep{krumholz_general_2005, padoan_star_2011}, which appear to agree well with numerical simulations and observations of molecular clouds \citep{federrath_star_2012}.  Computational experiments on the evolution of turbulence in isothermal ISM, considering the observed magnetic field distribution $B \propto \rho\gas^{1/2}$ \citep{crutcher_magnetic_1999, beig_observations_2003}, have additionally revealed more precise relationships between gas logarithmic density variance, sonic Mach number, and magnetic field strength \citep{molina_density_2012}.  However, the possible effects of magnetic fields and magnetised turbulent ISM models on galaxy evolution and properties have yet to be fully explored. MHD simulations of galaxy formation have thus far demonstrated that magnetic fields amplify rapidly within the turbulent rotating disk over the first $\sim 500~$Myr, before reaching a point of saturation and self-regulation. Their force then contributes to lower late-time star formation rates, relative to those in hydrodynamical disk galaxy simulations \citep{wang_magnetohydrodynamic_2009, dubois_magnetised_2010, pakmor_simulations_2013, steinwandel_origin_2020}. It remains an open question exactly what mechanisms drive this magnetic field evolution from its theorised weak primordial state to the $\gtrsim \mu\rm G$ field strengths observed today \citep{beck_magnetic_2016, han_observing_2017}, though numerical methods have been employed to probe a range of different possible dynamos: namely the galactic $\alpha$\textemdash$\Omega$ dynamo \citep{wang_magnetohydrodynamic_2009, dubois_magnetised_2010}, turbulent feedback driven dynamo \citep{pakmor_simulations_2013, vazza_amplification_2014, rieder_small-scale_2016}, magneto-rotational instability driven dynamo \citep{gressel_towards_2013, machida_dynamo_2013}, and cosmic-ray driven dynamo \citep{lesch_strong_2003, hanasz_global_2009}.  One persistent challenge in galaxy simulations to successfully model both this magnetised dynamo amplification and its effects on local galactic processes, as well as global galaxy evolution, is due to the dramatic spatial range these physics involve. In the case of star formation, certain studies opt to exclude it in their simulations in order to more rigorously model magnetic fields and the entire galaxy \citep[e.g.][]{wang_magnetohydrodynamic_2009, steinwandel_small-scale_2022}.  Others instead base their star formation routines on purely hydrodynamical methods, that then do not directly take into account local magnetic field properties \citep[e.g.][]{cen_galaxy_1992, springel_cosmological_2003}. \citet{katz_introducing_2021} examine primordial magnetic fields (PMFs) in a cosmomlogical context through a series of multi-frequency radiation-MHD simulations, and while their findings offer constraints on how PMFs may impact galaxy formation and evolution, the large spatial scales of their simulations prevent any detailed study of star-forming regions.  

To date, very few computational experiments exist that incorporate MHD-based star formation models into isolated disk galaxy simulations or simulations within a cosmological context. \citet{martin-alvarez_how_2020} have extended thermo-turbulent star formation models by \citet{trebitsch_fluctuating_2017}, \citet{mitchell_gas_2018} and \citet{rosdahl_snap_2017} into a magneto-thermo-turbulent (MTT) model, which they implement into a cosmological disk galaxy simulation. Their recipe identifies regions of gas collapse through defining an MTT Jeans length $\lambda_{\rm J,~MTT}$ dependent in part on the ratio of thermal to magnetic pressure, such that when the length of the gas cell surpasses $\lambda_{\rm J,~MTT}$ a certain mass fraction of gas is converted into stellar particles.  This fraction is determined by a local $\epsilon_{\rm ff}$, computed via the multi-scale model of \citet{padoan_star_2011}.  However the method of computing star formation efficiency remains dependent on solely hydro-dynamical parameters.  Additionally, this work along with others \citep[e.g.][]{vazza_amplification_2014, rieder_small-scale_2017} suffer from unrealistically slow magnetic field dynamos, due to the effects of limited spatial resolution on turbulent dissipation. There is still great need for a magnetised star formation recipe that can fully integrate the impact of magnetic fields on star formation efficiency, ideally with respect to new and improved models of generally unresolved small-scale turbulent dynamo.

In this work, we develop a new MHD-based star formation model that operates on sub-grid scales, deriving the local star formation efficiency $\eff$ from the properties of the sub-grid magnetised ISM turbulence.  We then use this star formation model in MHD simulations of isolated disk galaxies using the code \texttt{RAMSES} \citep{teyssier_cosmological_2002}. Our aim is to study the regulating effects of magnetic fields on star formation, identify how our magnetised model might impact the properties of galaxies on larger scales, and explain certain features observed in star-forming galaxies. 

In Section \ref{sec:model}, we derive our MHD sub-grid scale (SGS) star formation model from the pure hydrodynamics case previously developed by \citet{kretschmer_forming_2020}. We present in Section \ref{sec:methods} the numerical methods used in this paper, detailing additional SGS recipes for turbulence and stellar feedback.  In Section \ref{sec:results}, we describe the resulting galaxy disk morphology, magnetic field topology, and star formation history of our simulations. These properties and their implications are discussed in Section \ref{sec:discussion}.

\begin{figure*}
    \centering
    \includegraphics[width=0.9\textwidth]{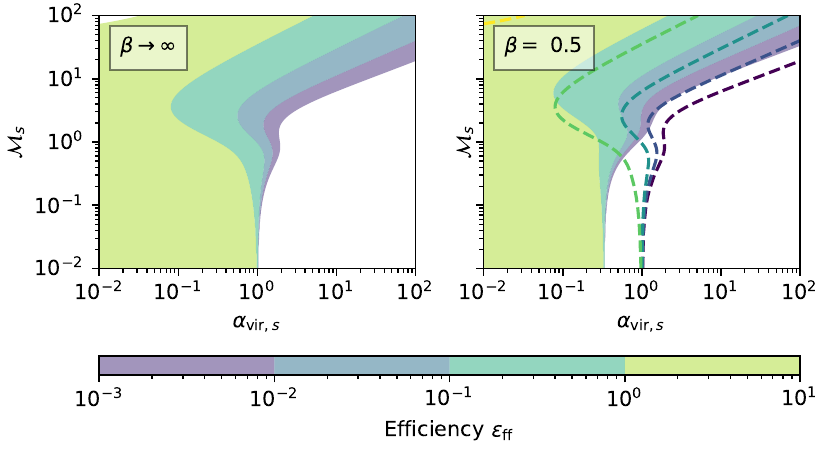}
    \caption{Contours of the efficiency per free-fall time $\epsilon_{\rm ff}$, computed for various combinations of sonic virial parameter $\alpha_{\mathrm{vir},~s}$ and sonic Mach number $\Ms$.  The left panel displays the shape of these contours in the non-magnetised sub-grid star formation recipe by \citet{kretschmer_forming_2020}, while the right panel displays their shape as per our magnetised model in the case of $\beta = 0.5$ ($\delta=3$).  At the low $\Ms$ limit, star formation efficiency depends on the entire sonic virial parameter of the cell, occuring at $\alpha_{\mathrm{vir},~s} \sim 1$ for the non-magnetised model and $\alpha_{\mathrm{vir},~f} \sim \delta \alpha_{\mathrm{vir},~s} \sim 1$ for the magnetised model. For $\Ms \gtrsim 1$ we see efficiency increasing in both models as $\alpha_{\mathrm{vir},~s}$ decreases, though the efficiency is generally dampened in the magnetised model as seen in the shift leftwards of efficiency contours.}
    \label{fig:eff_theory}
\end{figure*}

\section{Sub-grid scale star formation model}\label{sec:model}

Our star formation model draws upon that of \citet{kretschmer_forming_2020}, which requires the knowledge of the unresolved turbulent kinetic energy in each cell. This can be obtained using the so-called Large Eddy Simulations (LES) framework \citep[see e.g.][for relevant equations]{schmidt_numerical_2014, semenov_nonuniversal_2016}. In this framework a new hydro variable ($K_{\rm T}$) is introduced to represent the sub-grid turbulent kinetic energy, together with a new equation describing its evolution with proper source and sink terms according to mixing length theory \citep[see][for details]{kretschmer_forming_2020}. We can then recover the one-dimensional velocity dispersion of turbulence from this new variable at the scale of the mixing length, chosen here equal to the size of the cell $\Delta x$:
\begin{equation}
    K_{\rm T}= \frac{3}{2}\rho \sigma_{\rm T}^2
\end{equation}
This allows us to prolong the properties of turbulence down to unresolved scales $\ell \ll \Delta x$ using the classical scaling for supersonic Burgers turbulence with
\begin{equation}
    \sigma_{\rm 1D}(\ell) = \sigma_{\rm T} \left( \frac{\ell}{\Delta x} \right)^{1/2} \label{eq:Burger}
\end{equation}
We consider the fast magnetosonic length $\lf$, defined as the scale at which the 1D velocity dispersion $\sigma_{\rm 1D}(\lf)$ is equal to the fast magnetosonic speed,
\begin{equation}
    \cf \equiv \sqrt{\vA^2 + \cs^2}
    \label{eq:fastspeed}
\end{equation}
where $\vA = B/\sqrt{4\pi \rho}$ is the Alfv\'en speed, and $\cs$ is the sound speed. This is in essence the MHD counterpart of the sonic length, the length scale at which turbulence becomes transonic. A critical assumption in this definition is that $\vA$ is constant at sub-grid scales.  This is justified by simulations of supersonic and super-Alf\'venic turbulence in molecular clouds, in which $\vA$ approaches a mean value that is independent of density and corresponds to a power-law relation $B \propto \rho^{-0.4}$ \citep{padoan_superalfvenic_1999}. More recent simulations of supersonic magnetised turbulence performed by \citet{padoan_star_2011} have shown that $\vA$ is nearly constant within molecular clouds for densities $\rho \gtrsim 2\rho_0$, where $\rho_0$ is the mean density. Importantly, our assumption that $\vA$ is constant within our sub-grid model has no bearing on its properties at larger scales.  It remains an open question as to whether $\vA$ is constant throughout the galaxy.

Introducing the parameter $\beta = \cs^2 / \vA^2$, one can further relate the sound speed to the fast magnetosonic speed, 
\begin{equation}
    \cf = \sqrt{\cs^2 \left(\frac{1}{\beta}+1\right)} = \cs \sqrt{\frac{1+\beta}{\beta}}
\end{equation}
as well as the sonic Mach number of the turbulence $\Ms \equiv \sigma_{\rm 1D}/\cs$ to the fast magnetosonic Mach number,
\begin{equation}
    \Mf \equiv \frac{\sigma_{\rm 1D}}{\cf} = \Ms \sqrt{\frac{\beta}{1+\beta}}
\end{equation}
For brevity, we introduce from here on the magnetic correction factor $\delta \equiv (1+\beta)/\beta$, such that $\cf = \cs\sqrt{\delta}$ and $\Mf = \Ms / \sqrt{\delta}$.

Following for example \citet{federrath_star_2012}, we assume that the unresolved gas density distribution is described by a log normal PDF. Note that the use of a log normal PDF is justified by the multiplicative process through which density is distributed: shocks in the isothermal supersonic turbulent ISM are amplified via collisions and interactions with other shocks \citep{vazquez-semadeni_hierarchical_1994, passot_density_1998, kritsuk_statistics_2007, federrath_comparing_2010}.  This corresponds to an additive process in log space, which by the central limit theorem produces a Gaussian distribution for logarithmic density.

Defining $s = \ln(\rho/\bar{\rho})$ as the logarithmic density, where $\rho$ is the unresolved gas density and $\bar{\rho}$ is the mean gas density of the cell, this PDF takes the form
\begin{equation}
    p(s) = \frac{1}{\sqrt{2\pi\sigma_{s}^2}}\exp\left[-\frac{(s-\bar{s})^2}{2\sigma_{s}^2}\right]
    \label{eq:PDF}
\end{equation}
where $\bar{s}$ as the mean logarithmic density and $\sigma_{s}$ is the standard deviation. \citet{molina_density_2012} derive a dependence between $\sigma_{s}$ and $\Mf$ as
\begin{equation}
\sigma_{s}^2 = \ln(1 + b^2\Mf^2) = \ln\left(1 + \frac{b^2\Ms^2}{\delta}\right) 
\label{eq:sigMHD}
\end{equation}
where the parameter $b$ depends on the nature (compressive or solenoidal) of the turbulence forcing in the medium. It is evident here that the stronger the magnetic field, the narrower the variance of the logarithmic density PDF $\sigma_s^2$ will be, leading to fewer higher gas density regions.  This is further illustrated in simulations of MHD turbulence performed by \citet{federrath_star_2012}, where they found that as magnetic field strength increased, density contrast decreased and fewer fragments were formed overall. We recognise that while our model assumes $\sigma_s^2$ is isotropic, this does not necessarily hold for clouds with a trans or sub-Alfv\'enic magnetic field (where $\Ma \equiv \sigma_{\rm 1D}/\vA \lesssim 1$).  We hope that future work may capture such anisotropic regimes by incorporating density variance models that take into account preferential directions of density fluctuations \citep[e.g.][]{beattie_multi-shock_2021}

We now assume that our resolution element is described by an ensemble of small uniform clouds of diameter equal to the fast magnetosonic length $\lf$ and whose constant density follows the lognormal PDF introduced just above.  Following \citet{krumholz_general_2005}, we identify star-forming regions as the population of these small clouds for which the virial parameter is less than one. Using the classical spherical transonic cloud model of \citet{bertoldi_pressure-confined_1992}, for a cloud mass $M=4\pi/3 \rho R^3$ and a cloud radius $R=\lf/2$, we have
\begin{equation}
\alpha_f = \frac{5 \sigma_{\mathrm{tot}} R}{GM} = \frac{5(\sigma(\lf) + \cf^2)(\lf / 2)}{G (4\pi / 3) (\lf/2)^3 \rho}
    = \frac{15}{\pi} \frac{2\cf^2}{G \rho l_f^2} 
\end{equation}
We can simplify this expression using the fast magnetosonic Mach number as
\begin{equation}
\alpha_f = \frac{15}{\pi} \frac{2 \cf^2 M_f^4}{G \rho \Delta x^2} 
\label{eq:vir}
\end{equation}
We can also express this MHD-based virial parameter as a function of the old sonic length virial parameter $\alpha_s$ from \cite{kretschmer_forming_2020} as $\alpha_f = \alpha_s / \delta$.
Given that these clouds will collapse and form stars only if the virial parameter is smaller than one, this defines a critical collapse density at $\alpha_f = 1$:
\begin{equation*}
    \rho_{\mathrm{crit}} = \frac{15}{\pi}\frac{2 \cs^2 \Ms^4}{G \Delta x^2 \delta} \label{eq:rhocrit}
\end{equation*}
This critical density can be expressed in terms of the mean density $\bar{\rho}$ and the virial parameter of the cell $\alpha_{\mathrm{vir},f}$ as
\begin{equation}
    \rho_{\mathrm{crit}} = \alpha_{\mathrm{vir},f}\bar{\rho} \frac{2\Ms^4}{\delta(\delta + \Ms^2)}
\end{equation}
where
\begin{equation}
    \alpha_{\mathrm{vir},f} = \frac{15}{\pi} \frac{\sigma_{\rm 1D}^2 + \cf^2}{G \bar{\rho} \Delta x^2} = \frac{15}{\pi} \frac{\cs^2(\delta + \Ms^2)}{G \bar{\rho} \Delta x^2}
\end{equation}
Therefore the critical logarithmic density can be calculated as
\begin{equation}
    s_{\mathrm{crit}} = \ln\left(\frac{\rho_{\mathrm{crit}}}{\bar{\rho}}\right) = \ln \left[ \alpha_{\mathrm{vir},f} \frac{2\Ms^4}{\delta(\delta + \Ms^2)} \right] \label{eq:scrit}
\end{equation}
Note that this model breaks down at low Mach number \citep{federrath_star_2012}. Following \cite{kretschmer_forming_2020}, we modify Equation \ref{eq:scrit} such that for $\Mf \leq 1$, the lognormal PDF becomes a delta function and the stability criterion must be applied to the cell as a whole.  Just as in \citet{kretschmer_forming_2020}, this requires re-defining the critical logarithmic density as
\begin{equation}
    s_{\mathrm{crit}} = \ln \left[ \alpha_{\mathrm{vir},f} \left( 1 + \frac{2\Ms^4}{\delta(\delta + \Ms^2) }\right) \right] \label{eq:scrit-2}
\end{equation}
With the critical logarithmic density now properly defined in the general case, we can compute the local star formation efficiency using the multi-free fall time approach of \cite{federrath_star_2012},
\begin{align}
    \epsilon_{\mathrm{ff}} &= \int_{s_{\mathrm{crit}}}^{\infty} \frac{t_{\mathrm{ff}}(\bar{\rho})}{t_{\mathrm{ff}}(\rho)}\frac{\bar{\rho}}{\rho}p(s)\dd{s} \nonumber \\
    &= \frac{1}{2}\exp \left( \frac{3}{8}\sigma_{s}^2\right) \left[ 1 + \mathrm{erf}\left( \frac{\sigma_{s}^2 - s_{\mathrm{crit}}}{\sqrt{2\sigma_{s}^2}} \right) \right] \label{eq:eff}
\end{align}

Figure \ref{fig:eff_theory} displays the differences in star formation efficiency between the original star formation model of \citet{kretschmer_forming_2020} and our MHD version for a typical magnetised case with $\beta=0.5$. While the efficiency contours have similar shapes, the magnetic field effectively reduces the star formation efficiency in the parameter space defined by the cell sonic virial parameter and the sub-grid turbulence sonic Mach number.  Note that 
\begin{equation}
    \alpha_{\mathrm{vir},f} = \alpha_{\mathrm{vir},s}\left(\frac{\delta + \Ms^2}{1 + \Ms^2} \right)
\end{equation}
where $\alpha_{\mathrm{vir},s}$ is the virial parameter of the whole cell in the non-magnetised model. The sonic virial parameter emerges in the non-magnetised limit, $\beta \rightarrow \infty$ or $\delta \rightarrow 1$. The presence of magnetic fields in a potentially collapsing molecular core will increase the value of the virial parameter. Hence smaller effective sonic virial parameters are necessary for the magnetised model to produce the same efficiencies of star formation. In Figure \ref{fig:eff_theory}, this corresponds to a critical value of $\alpha_{\mathrm{vir},s} \sim 0.1$ in the magnetised case with $\beta = 0.5$, versus  $\alpha_{\mathrm{vir},s} \sim 1$ in the non-magnetised case. 

We see in both models the dramatic break in efficiency at low sonic Mach numbers, as in this regime both models allow for star formation via a single threshold virial parameter.  This threshold differs by a factor of $1/\delta$, which corresponds to the relation $\alpha_{\mathrm{vir},s}=\alpha_{\mathrm{vir},f} / \delta$ in the low $\Ms$ limit. At larger sonic Mach numbers, the virial parameters $\alpha_{\mathrm{vir}, s}$ and $\alpha_{\mathrm{vir}, f}$ approach equality but the critical logarithmic density increases in the magnetised model relative to the non-magnetised model: $s_{\rm crit} \approx \ln\left[ \alpha_{\mathrm{vir}, s} (1 + \frac{2}{\delta}\Ms^2) \right]$. As such, efficiency is suppressed when magnetic fields are present, as lower effective sonic virial parameters are required in the magnetised model to reach the same efficiencies as before.  This is illustrated by the slight translation leftwards in the efficiency contours between the two plots. 

\section{Numerical methods}\label{sec:methods}

To compare our new star formation model to previous models, we have performed a suite of  isolated disk galaxy simulations with the octree-based Adaptive Mesh Refinement (AMR) code \texttt{RAMSES} \citep{teyssier_cosmological_2002}. In these simulations, dark matter and stars are modelled as a collisionless fluid, while the gas component follows the ideal MHD equations (see below). These different fluids are coupled through self-gravity. 

Our initial conditions are taken from the high-resolution \textit{AGORA} simulation project \citep[see][]{kim_agora_2016}. They correspond to a typical Milky-Way-sized, gas-rich disk galaxy embedded in its parent dark halo. The simulation is initialised with 5 different components: the dark matter halo, the gaseous halo, the stellar disk, the stellar bulge, and the gaseous disk. We use the DICE code to generate our initial particle and gas cell  distributions using a Metropolis Hasting Monte Carlo Markov Chain algorithm \citep[see details in ][]{perret_dice_2016}.

The dark matter halo is distributed according to a \citet{navarro_universal_1997} profile with concentration parameter $c=10$, spin parameter $\lambda = 0.04$, $M_{200} = 1.074 \times 10^{12}~M_{\odot}$, $R_{200}=205.5$ kpc, and circular velocity $v_{\mathrm{c}, 200} = 150$ km s$^{-1}$.  

The gaseous halo is filled with a very low uniform density of $n_{\mathrm{H}} = 10^{-6}$ cm$^{-3}$ and a uniform temperature of $10^6$ K, zero initial velocity, and zero metallicity.  

The disk component contains a total mass of $M_{\mathrm{d}} = 4.297 \times 10^{10}~M_{\odot}$ with scale length $r_{\mathrm{d}} = 3.432$ kpc, scale height $z_{\mathrm{d}} = 0.1r_{\mathrm{d}}$, gas mass fraction $f_{\mathrm{gas}} = M_{\mathrm{d},\mathrm{gas}} / M_{\mathrm{d}}=0.2$, which corresponds to a disk gas-to-star ratio of $1/4$.  

Both the stellar and gas component of the disk follow an exponential density profile
\begin{equation}
\rho_{\mathrm{d},i}(r,z) = \rho_{0,i} ~e^{-r/r_{\mathrm{d}}} \cdot e^{-|z|/z_{\mathrm{d}}}
\end{equation}
where index $i$ stands here  either for gas or stars, $r$ is the cylindrical radius, and $z$ is the vertical height from the disk plane.
The central disk densities are computed using 
\begin{equation}
\rho_{0,\mathrm{i}} = \frac{M_{\mathrm{d},\mathrm{i}}}{4\pi r_{\mathrm{d}}^2 z_{\mathrm{d}}}
\end{equation}
Both gas and stellar disks are truncated at $3 r_{\rm d}$ and $3 z_{\rm d}$. The stellar bulge is modelled with a \citet{hernquist_analytical_1990} density profile and stellar mass $M_{\mathrm{b},*} = 4.297 \times 10^9~M_{\odot}$. 

Our initial gas disk is turbulent. We add random velocity fluctuations on top of the overall rotating flow. These fluctuations follow a Burgers $k^{-2}$ kinetic energy spectrum normalized to $\sigma_{\rm 1D}=20$~km/s. We found this to be a crucial addition compared to previous work \citep[see][]{kim_agora_2016}. Indeed, a disk in perfect hydrostatic equilibrium would collapse vertically as it cools. The resulting razor-thin gas slab will trigger an unrealistically strong starburst. 

Our new star formation model is quite sensitive to the magnitude of the sub-grid turbulence. We initialise the sub-grid turbulence to the stationary value obtained by balancing source terms and sink terms in the LES sub-grid kinetic energy equation \citep{kretschmer_forming_2020}. These precautions explain why our star formation history is so smooth and slowly declines in time without a spurious prominent starburst at startup (see the Results section).

We are particularly interested in the evolution of the magnetic field and its impact on star formation. The outcome of the simulation will depend on the initial field strength and topology of the field. In this paper, we restrict ourselves to a very weak initial field, so that the final field strength and topology are solely a consequence of our galactic dynamo. We will study the impact of the initial (or fossil) magnetic field in a follow-up paper. The smallest value one can realistically consider for the initial field is the one produced by the Biermann battery during the linear regime of cosmological density perturbations \citep{naoz_generation_2013}.  We initialise our magnetic field with a constant magnitude of 10$^{-20}$~G and parallel to the $z$ direction.

\subsection{Details on the adaptive grid}

Simulations begin from a uniform grid covering our entire computational box of size (320 kpc)$^3$, with a coarse grid of resolution 128$^3$ corresponding to our coarsest refinement level $\ell_{\mathrm{min}}=7$ and a coarsest cell size of 2.5 kpc. Our finest refinement level was set to $\ell_{\mathrm{max}}=14$ so that our finest resolution element size is $\sim 20$~pc. We employ a quasi-Lagrangian strategy: we trigger a new cell refinement if the baryonic mass (gas plus stars) within the cell exceeds $10^3~M_{\odot}$, or if the cell contains more than 8 collisionless particles.  Boundary conditions are set to isolated for the gravity solver and out-flowing for the MHD solver.

\subsection{Details on the MHD and gravity solvers}

We give her more details on our MHD solver, as it entails the core of our study. We solve the following ideal MHD equations, assuming that the gas is ionised enough to justify neglecting all non-ideal MHD effects. Written in conservative form, these equations are
\begin{align}
	\partial_t \rho + \nabla \cdot (\rho \mathbf{v}) &= 0\\
	\partial_t(\rho \mathbf{v}) + \nabla \cdot (\rho\mathbf{v}\mathbf{v}^{\mathrm{T}} - \mathbf{B}\mathbf{B}^{\mathrm{T}} + P_{\mathrm{tot}}) &= \rho \mathbf{g}\\
	\partial_t E + \nabla \cdot [(E+P_{\mathrm{tot}})\mathbf{v} - (\mathbf{v} \cdot \mathbf{B})\mathbf{B}] &= \Gamma - \Lambda + \rho \mathbf{g} \cdot \mathbf{v} \\
	\partial_t \mathbf{B} - \nabla \times (\mathbf{v} \times \mathbf{B}) &= 0 \label{eq:induction-eq} \\
	P_{\mathrm{gas}} &= (\gamma - 1) \rho \epsilon
\end{align}
where $\rho$ is the mass density, $\rho\mathbf{v}$ is the momentum, $\mathbf{B}$ is the magnetic field, $\mathbf{g}$ is acceleration due to gravity, $E = \frac{1}{2}\rho\mathbf{v}^2 + \rho\epsilon + \frac{1}{2}\mathbf{B}^2$ is the total fluid energy, $P_{\mathrm{tot}} = P_{\mathrm{gas}} + \frac{1}{2}\mathbf{B}^2$ is the total pressure. In the energy equation, $\Gamma$ is the heating function and $\Lambda$ is the cooling function, $P_{\mathrm{gas}}$ is the pressure of the gas, and $\epsilon$ is the specific internal energy. We consider an ideal gas equation of state with adiabatic exponent $\gamma = 5/3$. 

Our numerical scheme is a standard second-order Godunov scheme with the MUSCL-Hancock scheme coupled to the Constrained Transport method for the induction equation \citep{teyssier_kinematic_2006}. The CT scheme enforces the solenoidal constraint $\nabla \cdot {\bf B}=0$ to machine precision accuracy. As a result, our numerical scheme conserves mass, linear momentum and total fluid energy, as well as magnetic flux. 

As explained in \citet{fromang_high_2006}, we use the HLLD Riemann solver to compute intercell fluxes of mass, momentum and energy. We use a 2D version of the HLLD Riemann solver to compute the electric field at the edge of the faces where the magnetic field is defined \citep{fromang_high_2006}. Second-order reconstruction of primitive variables is performed using the \textit{MinMod} slope limiter \citep[see e.g.][]{roe_characteristic-based_1986}. When new cells are refined, we use straight injection of conservative variables from the parent cell. Last but not least, we use for self-gravity the Multigrid Poisson solver available in \texttt{RAMSES} \citep{guillet_simple_2011}. 

\begin{figure*}
\includegraphics[width=\linewidth]{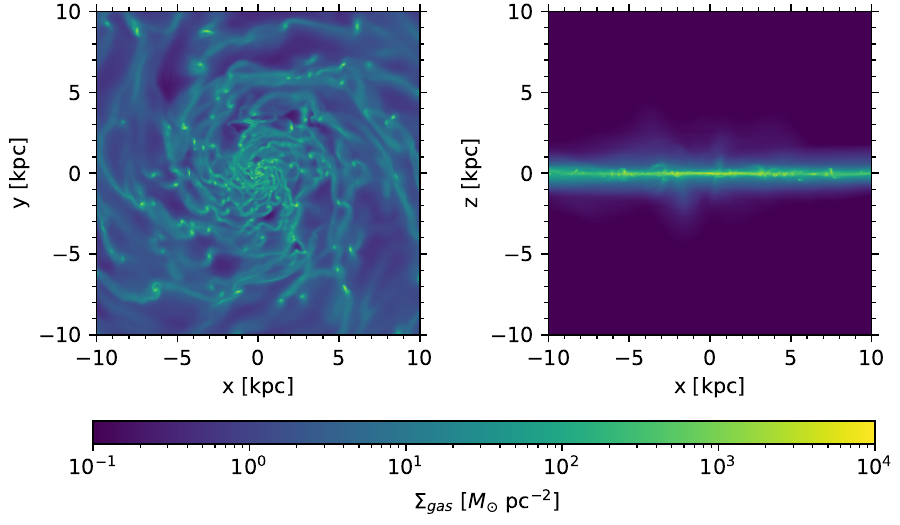}\\
\includegraphics[width=\linewidth]{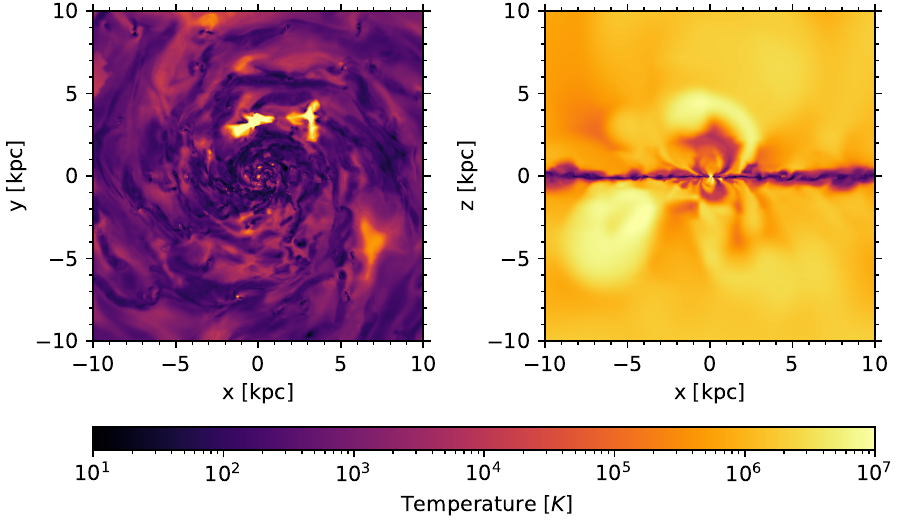}
\caption{Face-on and edge-on maps of gas surface density $\Sigma\gas$ (top row) and temperature (bottom row) of our simulated galaxy.  Both quantities are density-weighted, and the gas surface density is averaged along the line of sight within a box 320 kpc wide in each direction. For the temperature map, face-on and edge-on views are taken from the $z=0$ and $y=0$ plane respectively.}
\label{fig:dens_temp_maps}
\end{figure*}

\subsection{Details on the sub-grid physics}

In this paper, we compare two simulations with two different star formation models. The first employs, as a reference, the non-magnetised sub-grid star formation recipe developed originally in \citet{kretschmer_forming_2020}.  The second uses our new sub-grid star formation model, in which the effects of magnetic fields on the local efficiency are considered (see previous section). Both simulations incorporate the LES model for turbulent kinetic energy, providing us with the sub-grid turbulent kinetic energy that enters our sub-grid star formation models. 

We also use the same turbulent kinetic energy information for our implementation of a sub-grid magnetic dynamo, as described  briefly below. Beyond these important sub-grid models, we have used the traditional galaxy formation  ingredients provided by \texttt{RAMSES} code, such as equilibrium H and He gas cooling \citep[e.g.][]{katz_damped_1996} plus metal cooling at both low and high temperature \citep{sutherland_cooling_1993}, heating by a standard  \citet{haardt_radiative_1996} UV background and a self-shielding density of $n_{\rm H} = 0.01$~cm$^{-3}$  \citep{aubert_reionization_2010}.

\subsubsection{Sub-grid Turbulent Dynamo}

As in \citet{liu_subgrid_2022}, we model the impact of unresolved turbulence on the magnetic field evolution using a mean-field approach \citep[see e.g.][and references therein]{schmidt_fluid-dynamical_2011}.  Assuming homogeneity and isotropy of the turbulence, one can derive a tensor $\alpha$ relating the mean magnetic field $\mathbf{\bar{B}}$ to the turbulent electromotive force caused solely by unresolved turbulent fluctuations in the velocity and magnetic fields, $\bar{\mathcal{E}}$: 
\[\bar{\mathcal{E}} = \alpha \mathbf{\bar{B}}\] 
A simple model for $\alpha$ has been proposed by \citet{liu_subgrid_2022} that depends on the local sub-grid turbulent velocity dispersion $\sigma_{\rm 1D}$ 
\[\alpha = \sigma_{\rm 1D} \mathrm{max}\left[ 1 - \frac{E_{\mathrm{mag}}}{qK_{\mathrm{T}}}, 0 \right]{\mathbb I}\]
where $E_{\mathrm{mag}} = \bar{B}^2/8\pi$ is the mean-field magnetic energy density, $q$ is a turbulent dynamo quenching parameter, and $K_{\mathrm{T}} = \frac{3}{2}\rho \sigma_{\rm 1D}^2$ is the sub-grid turbulent kinetic energy density. 

For these simulations, we adopt the fiducial quenching parameter $q = 10^{-3}$ \citep{liu_subgrid_2022}. This means that our sub-grid turbulent dynamo will vanish entirely as soon as the magnetic energy reaches 0.1\% of the local turbulent energy. For typical gas disk conditions with $n_{\rm H} \simeq 1$~cm$^{-3}$ and $\sigma_{\rm 1D} \simeq 1-10$~km/s, this corresponds to a mean field of roughly $0.02-0.2$~$\mu$G.

We also limit the sub-grid turbulent dynamo model to only cooling and star-forming gas, by setting $\alpha = 0$ for regions with number densities less than the self-shielding threshold $n_{H} = 10^{-2}$ cm$^{-3}$. In this manner, strong magnetic fields in the circumgalactic medium can only arise from galactic outflows and not from a local turbulent dynamo, unless it is resolved by the code, which is unlikely given our limited resolution. Previous works have shown that when the resolution is high enough, dynamo amplification could emerge from resolved motions without relying on a sub-grid turbulent dynamo model \citep[see][]{rieder_small-scale_2016, vazza_resolved_2018,steinwandel_small-scale_2022}.

\begin{figure*}
    \centering
    \includegraphics[width=0.9\textwidth]{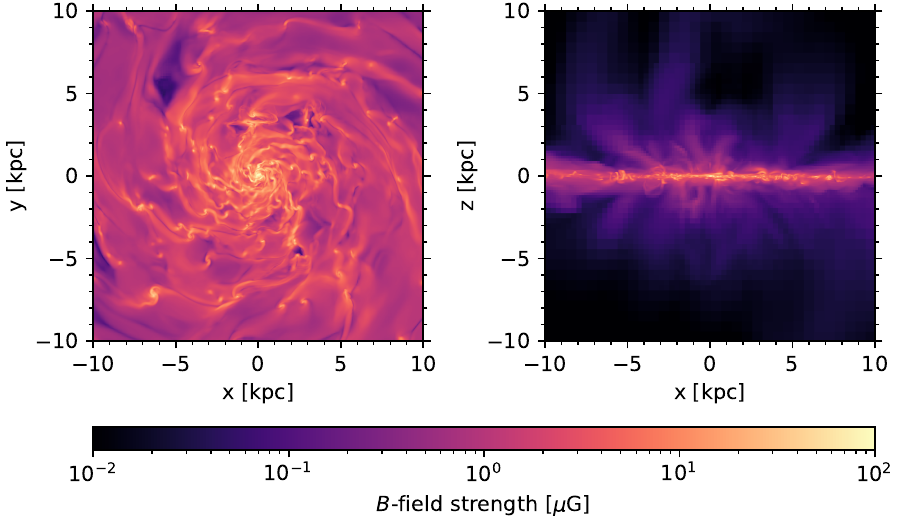}
    \caption{Face-on and edge-on views of the density-weighted magnetic field strength, within the $z=0$ plane (left) and $y=0$ plane (right).}
    \label{fig:B_map}
\end{figure*}

\subsubsection{Sub-grid Stellar feedback}

Our stellar feedback model follows that of \citet{kretschmer_forming_2020}, which quantifies for each stellar particle (with typical mass $10^5~M_{\odot}$) the scalar momentum due to sub-grid supernovae explosions, and injects this momentum into the surrounding gas.  Assuming a uniform distribution of individual supernova (SN) explosion times for a given stellar particle between $t_{\mathrm{start}} = 3$ Myr and $t_{\mathrm{end}} = 20$ Myr, the rate of supernovae can be computed as
\begin{equation}
    \dot{N}_{\mathrm{SN}} = \frac{\eta_{\mathrm{SN}} m_{\rm ini,*}}{M_{\mathrm{SN}} (t_{\mathrm{end}} - t_{\mathrm{start}})} 
\end{equation}
where $\eta_{\mathrm{SN}}=0.2$ is the mass fraction of the stellar particle producing supernovae, $m_{\rm ini,*}$ is the initial stellar particle mass, and $M_{\mathrm{SN}}=10~M_{\odot}$ is the typical supernova progenitor mass. Hence the expected number of supernovae within a timestep $\delta t$ is
\begin{equation}
    \langle N_{\mathrm{SN}} \rangle = \dot{N}_{\mathrm{SN}} \delta t
\end{equation}
For each timestep, we can draw the actual number $N_{\mathrm{SN}}$ of supernovae for a stellar particle of mass $M_{i,*}$ from a Poisson distribution with average $\langle N_{\mathrm{SN}} \rangle$.  The minimum scale at which momentum must be injected is determined by the cooling radius $R_{\mathrm{cool}}$, which is modelled as
\begin{equation}
    R_{\mathrm{cool}} = 3.0~\mathrm{pc}~\left(\frac{Z}{Z_{\odot}} \right)^{-0.002} \left( \frac{n_{\mathrm{H}}}{100~\mathrm{cm}^{-3}} \right)^{-0.42}
\end{equation}
where $Z$ is the gas metallicity and $n_{\mathrm{H}}$ the number density of hydrogen. To account for the cases in which $R_{\mathrm{cool}}$ is either resolved or unresolved by at least one grid cell with length $\Delta x_{\mathrm{min}}$, we calculate the total resulting momentum as
\begin{align}
    P &= P_{\mathrm{SN}}N_{\mathrm{SN}}~ \mathrm{min}\left(1,\left(\frac{\Delta x_{\mathrm{min}}}{R_{\mathrm{cool}}}\right)^{3/2} \right)\\
    \intertext{where}
    P_{\mathrm{SN}} &= 1.42 \times 10^5~\mathrm{km~s}^{-1}M_{\odot}~\left(\frac{Z}{Z_{\odot}} \right)^{-0.137} \left( \frac{n_{\mathrm{H}}}{100~\mathrm{cm}^{-3}} \right)^{-0.16}
\end{align}
This momentum is then injected isotropically into the surrounding gas cells using the numerical techniques described in \citet{kretschmer_forming_2020}.

\begin{figure*}
\includegraphics[width=0.9\linewidth]{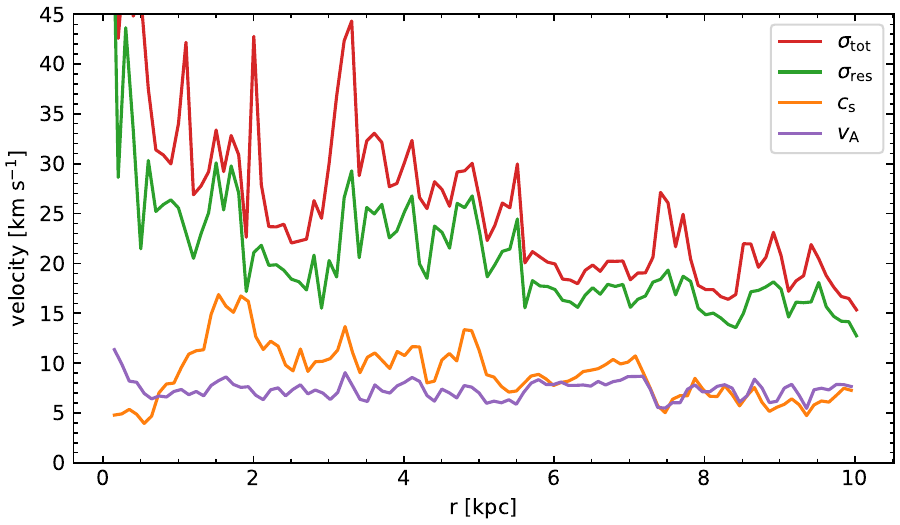}
\caption{Total resolved plus sub-grid velocity dispersion $\sigma_{\rm tot}$ (red line), resolved velocity dispersion $\sigma_{\rm res}$ (green line), sound speed $\cs$ (orange line), and Alfv\'en velocity $\vA$ (purple line) as varying with radius $r$ in the central region of the disk. The disk profile has a maximum radius of 10 kpc and height of 3.2 kpc. Quantities are computed as mass-weighted averages within 100 cylindrical bins.}
\label{fig:velocity_compare}
\end{figure*}

\begin{figure*}
    \centering
    \includegraphics[width=0.9\textwidth]{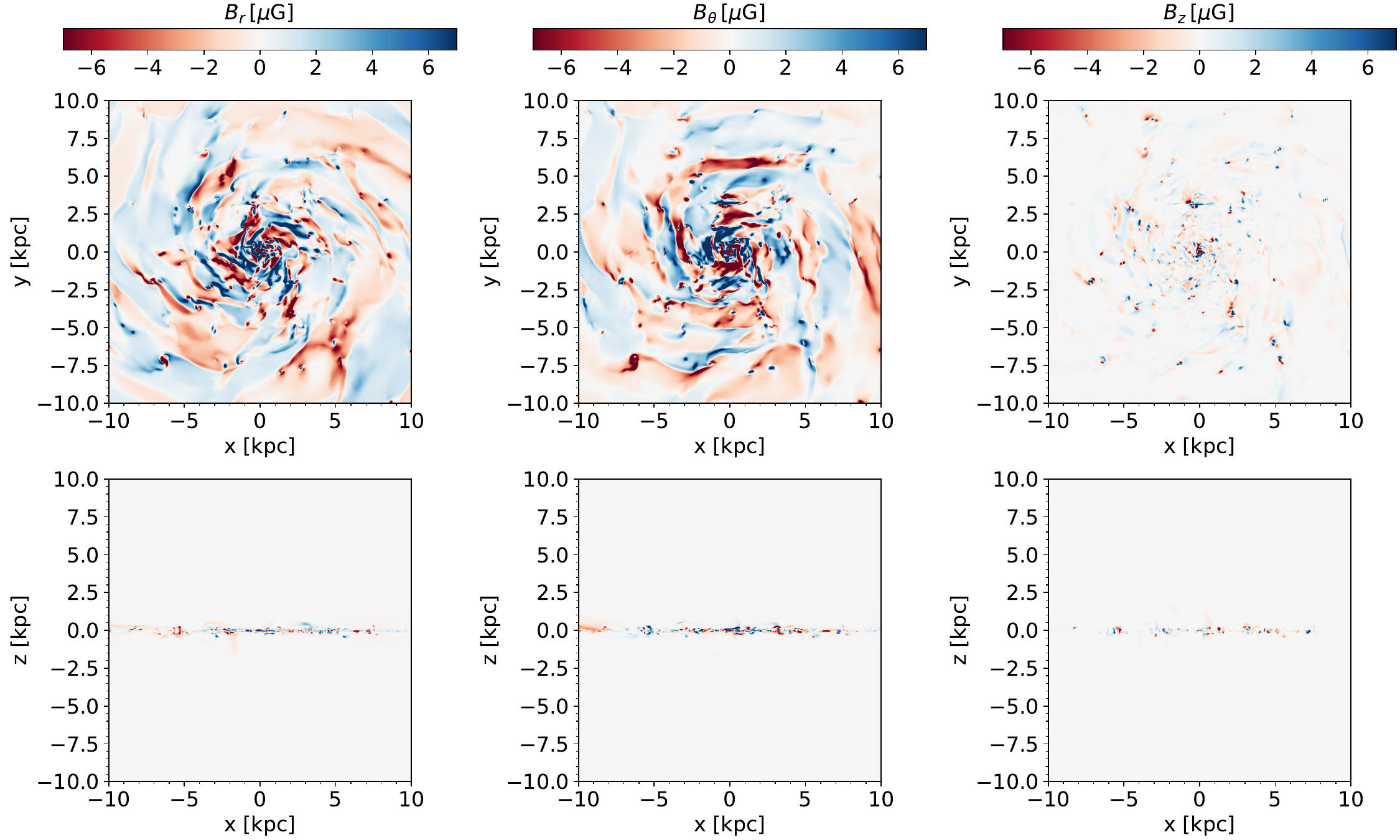}
    \caption{Radial ($B_r$, left column), toroidal ($B_\theta$, middle column), and vertical ($B_z$, right column) components of the density-weighted magnetic field, mapped either face-on (top row) or edge-on (bottom row) in the $z=0$ plane and $y=0$ plane.}
    \label{fig:Bcylmap}
\end{figure*}

\section{Simulation results}\label{sec:results}

In this section, we describe in detail our simulation results. Our final snapshot has reached a time of 1.4~Gyr, for which the galaxy is forming stars steadily at a rate of 1 solar mass per year. A galactic fountain is clearly visible and plays an important role in regulating the gas available for forming stars in the disk. After an initial phase of fast amplification due to our sub-grid mean field dynamo model, the magnetic field quickly saturates slightly below equipartition with the gas turbulent energy and develops a strong toroidal structure, although in our particular case, it appears to be still dominated by small scale fluctuations.

\subsection{Disk morphology}

\begin{figure*}
    \centering
    \includegraphics[width=0.9\textwidth]{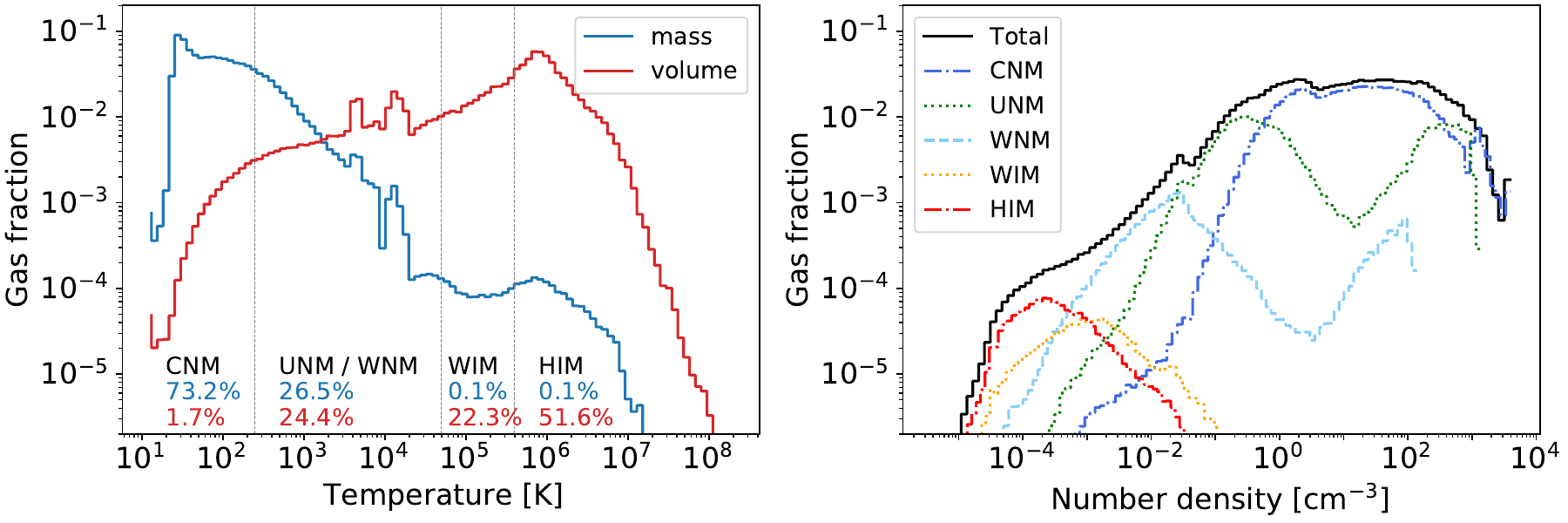}
    \caption{Left panel: mass-weighted (blue) and volume-weighted (red) gas fractions as a function of temperature, radially binned in 100 bins across a disk profile around the centre with maximum radius of 10 kpc and height of 3.2 kpc. Different ISM phase regions are annotated, along with the total mass-weighted (blue) and volume-weighted (red) mass fractions of each phase. They are the cold neutral medium (CNM; $T < 250~$K), the thermally unstable (UNM; $250~\mathrm{K} \leq T < 3 \times 10^3~$K) and warm neutral medium (WNM; $3 \times 10^3~\mathrm{K} \leq T < 5 \times 10^4~$K), the warm ionised medium (WIM; $5 \times 10^4~\mathrm{K} \leq T < 4 \times 10^5~$K), and the hot ionized medium (HIM; $4 \times 10^5~\mathrm{K} \leq T$). Right panel: Mass-weighted gas fraction as a function of number density over the same binned cylindrical region as the left panel. We show this density PDF for all of the gas in the region (black), the cold neutral medium phase (blue), the thermally unstable phase (green), the warm neutral medium phase (light blue), the warm ionised phase (orange), and the hot ionised phase (red).}
    \label{fig:gas_fractions}
\end{figure*}

Figure \ref{fig:dens_temp_maps} shows edge-on and face-on images of the gas surface density and the mass-weighted temperature of our simulated galaxy at $t=1.4$~Gyr.  The gas surface density reaches its maximum value around $10^3~M_{\odot}~$pc$^{-2}$ in the nuclear region, with multiple cold and dense star-forming gas clumps visible along the spiral arms.  We note that the apparent spiral structure of the gas clumps themselves, while differing from observed complex morphologies of such objects, is in fact expected.  The limited resolution of our simulations ($\sim$ 20 pc scale) prevents us from resolving the clouds' internal structure, hence these clouds cannot fragment into the necessary substructures.  However, the Coriolis force associated with galactic rotation still moulds clouds into a spiral shape.  We expect that, were our simulations able to resolve scales below typical gas cloud radii, the gas clouds of Fig. \ref{fig:dens_temp_maps} would fragment further into various dense clumps.  Such fragmentation is visible in sub-parsec resolution simulations of spiral galaxies, with smaller gas clumps roughly distributed along spiral arms like the "beads of a string" \citep[see e.g.][]{renaud_sub-parsec_2013}.

The multiphase nature of the interstellar gas (ISM) appears evident in Figure~\ref{fig:dens_temp_maps} with low density, hot supernova-driven bubbles ($T \simeq 10^7$~K) coexisting alongside intermediate density warm gas ($T \simeq 10^3-10^4$~K) and high density cold clumps ($T \simeq 10-100$~K). In the edge-on view, we can clearly see the galactic fountain as thin streams of gas are ejected outwards from and fall back towards the galactic plane. These processes heat the low density circumgalactic medium (CGM), which as seen in Figure \ref{fig:dens_temp_maps} ranges in temperature from $10^4-10^7~$K. Due to our simulation's limited resolution outside of the galactic disk, the multiphase structure of the CGM \textemdash\, which has been observed to comprise of not only hot virialised gas, but also atomic hydrogen (with temperatures $T\sim 10^4$ K) and molecular hydrogen ($T\sim 50$ K) in circumgalactic gas \citep[for review, see][]{tumlinson_circumgalactic_2017} \textemdash\, cannot be fully reproduced.  Given that our work is primarily focused on star formation within the disk, we leave more robust treatments of and refinement within the CGM for future computational experiments.

\subsection{Magnetic field topology}

\begin{figure*}
    \centering
    \includegraphics[width=0.9\textwidth]{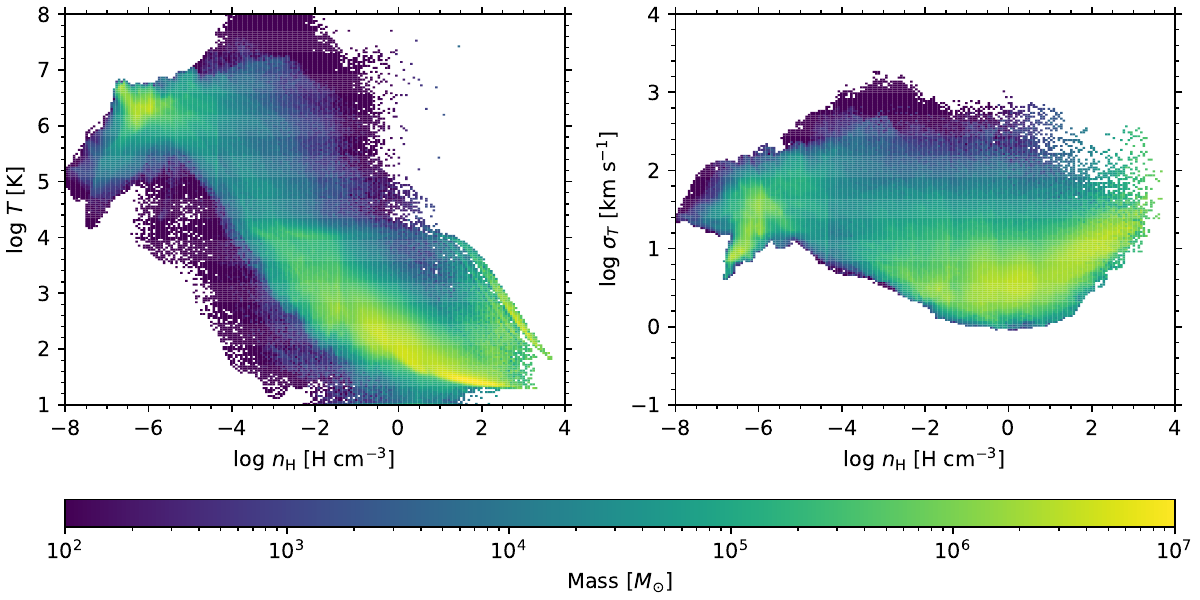}
    \caption{Mass-weighted histograms of temperature (left) and sub-grid velocity dispersion $\sigma_T$ (right) versus number density $n_{\rm H}$.}
    \label{fig:sigmatemp-vars-rho}
\end{figure*}

\begin{figure*}
    \centering
    \includegraphics[width=0.9\textwidth]{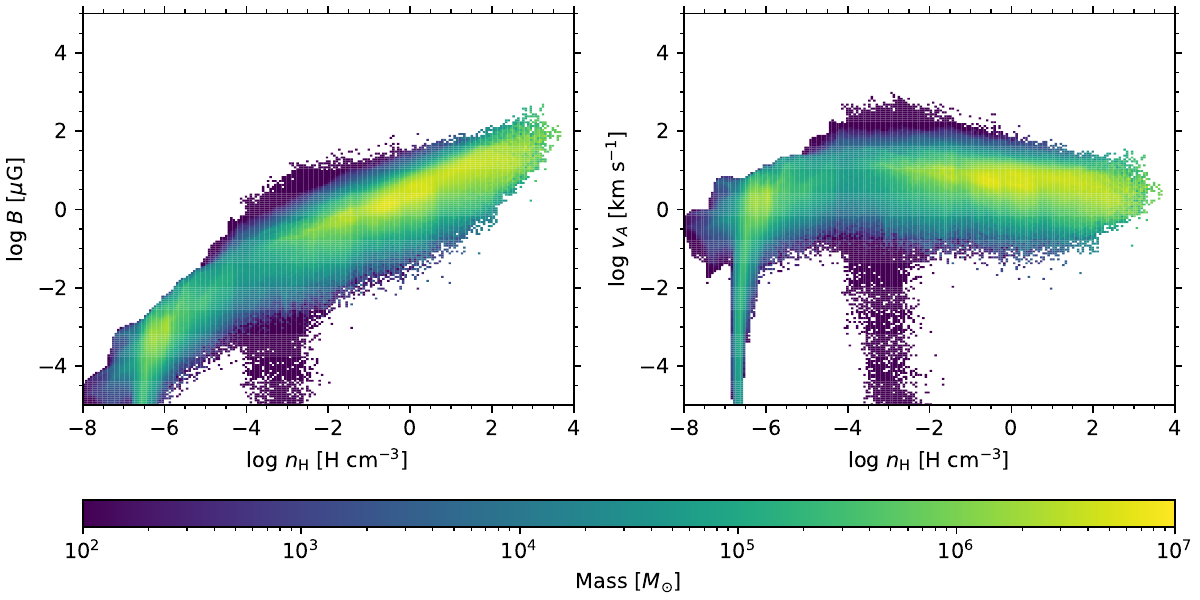}
    \caption{Mass-weighted histograms of  magnetic field strength and Alfv\'en velocity $\vA$ versus number density $n_{\rm H}$.}
    \label{fig:BvA-vars-rho}
\end{figure*}

After 1.4 Gyr, our simulated galaxy displays a relatively ordered magnetic field that closely follows the gas density (see Fig.~\ref{fig:B_map}). The central star-forming region has the highest magnetic field strengths around $100~\mu$G, while in the disk, dense clumps with low temperatures $T \lesssim 10^2~$K and strong magnetic fields $B \gtrsim 10~\mu$G accumulate along the spiral arms.  The magnetic field strength generally decreases with increasing radius. Our magnetic energy is roughly in equipartition with the thermal energy but not with the turbulent energy (see Fig.~\ref{fig:velocity_compare}). Moreover, when averaged over cylindrical bins, the magnetic energy is dominated by the fluctuating radial and tangential components, while the mean radial and tangential fields are almost zero. This is consistent with our very weak initial magnetic field and indicates that our magnetic field evolution converges towards a saturated small-scale turbulent dynamo.

In Figure \ref{fig:Bcylmap}, we have plotted the mass-weighted, radial, tangential and vertical components of the magnetic field, seen both face-on and edge-on. Although the toroidal component appears to be the strongest, we see also a strong radial component and a  weak vertical component. We obtain main field reversals as revealed by the multiple changes of sign of both the radial and the tangential component, explaining why the average values vanish. Along the spiral arms, $B_r \sim B_\theta \sim 5~\mu$G, while in the regions between spiral arms, the magnetic field component magnitudes slightly decrease to $\sim 2~\mu$G.

\subsection{Structure of the ISM}
We distinguish the various phases of our simulated galaxy ISM through the mass-weighted and volume-weighted density distributions of the gas, visualised as a function of temperature in the left panel of Figure \ref{fig:gas_fractions}. Cold, dense gas (which we demarcate via a temperature criterion of $T < 250~$K) clearly dominates at about 73\% of the mass fraction, consistent with our expectations for a star forming disk.  However, there is not a clear distinction between this cold neutral medium (CNM) and a warm neutral medium (WNM) in our ISM, aside from a local peak in the gas fraction that occurs around $T \sim 2 \times 10^4~$K.  In this final simulation snapshot the galaxy is still relatively gas-rich (with a total gas fraction $f_\mathrm{gas} \sim 20$\%) and actively star-forming, which suggests the presence of a thermally unstable neutral medium (UNM).  Furthermore, our limited numerical resolution prevents proper spatial resolution of this transitional medium between the CNM and WNM. Hence we consider in the left panel of this figure the UNM combined with the WNM, as opposed to their fractions individually, to account for gas within the temperature range of  $250~\mathrm{K} \leq T < 5 \times 10^4~$K. Gas at these temperatures make a non-negligible contribution to the mass fraction of $\sim 26$\%.  Additionally, we label an apparent warm ionised medium (WIM) with temperatures $5 \times 10^4~\mathrm{K} \leq T < 4 \times 10^5~$K, and a hot ionised medium (HIM) with temperatures $4 \times 10^5~\mathrm{K} \leq T$.  These phases make up the least of the gas mass, while being the most volume filling: the HIM fills 51.6\% of the volume, and the WIM fills 22.3\%.  The UNM/WNM phase also makes a non-neglible contribution to the volume with a 24\% volume-filling fraction, and the smallest (1.7\%) volume-filling fraction can be attributed to the CNM.  These values align with our expectations and previous results on the ISM structure of star forming, gas-rich disks \citep[e.g.][]{hopkins_structure_2012}.

{Differences in our denoted ISM phases are further illustrated in the right panel of Fig. \ref{fig:gas_fractions}, where we plot mass-weighted gas fraction as a function of number density.  Here we have differentiated the UNM from the WNM, using gas temperature cutoffs of $250~\mathrm{K} \geq T < 3 \times 10^3~$K and $3 \times 10^3~\mathrm{K} \geq T < 5 \times 10^4~$K for the UNM and WNM respectively.  The HIM and WIM correspond to lower density gas, with their probability density functions (PDFs) centred at number densities of $10^{-4}~\rm cm^{-3}$ and $10^{-3}~\rm cm^{-3}$ respectively.  The UNM and WNM lie largely in the expected intermediate density range of $0.01~\rm cm^{-3}$ \textemdash $1~\rm cm^{-3}$, while the CNM is concentrated at high densities ($>1~\rm cm^{-3}$).

Figure \ref{fig:sigmatemp-vars-rho} shows 2D mass-weighted histograms of temperature and sub-grid velocity dispersion as a function of gas density. Similar to Fig. \ref{fig:gas_fractions}, the temperature histogram reflects how the bulk of gas mass lies in a cold phase with number densities $\sim 100 \mathrm{cm}^{-3}$ and temperatures $\lesssim 50~$K.  Warmer, intermediately-dense gas ($T\sim 1000~$K, $n_{\rm H} \sim 0.1~\rm cm^{-3}$ makes up the second largest portion of the mass, while the most diffuse and hot gas ($T > 10^5~$K, $n_{\rm H} < 0.01~ \rm cm^{-3}$ makes up the least.  We note that a seemingly concentrated portion of mass at temperatures $T\sim 10^6~$K and densities $n_{\rm H} \sim 10^{-6}~\rm cm^{-3}$ are again numerical artefacts, attributable to the lack of resolution in our galaxy CGM.  The sub-grid velocity dispersion reaches its minimum around $\sigma_{\rm 1D}\sim 1~\mathrm{km}~\mathrm{s}^{-1}$ for a moderate density around $1 \mathrm{cm}^{-3}$, close to the average gas density in the disk. Inside dense clumps, however, the sub-grid velocity dispersion increases following the expected adiabatic relation $ \sigma_{\rm 1D} \propto \rho^{1/3}$ \citep{semenov_nonuniversal_2016}.

\subsection{Why is the Alfv\'{e}n speed so constant?}

The 2D mass-weighted histograms of magnetic field strength and Alfv\'{e}n speed as a function of gas density, shown in Figure \ref{fig:BvA-vars-rho}, reflect typical magnetic field strengths and densities in the disk on the order of $10~\mu$G and $1 \mathrm{cm}^{-3}$, respectively. Maximum strength of $100~\mu$G is particularly evident in the densest regions, namely cold clumps along the spiral arms and the nuclear region of the disk.  A striking result visible in this figure is that the Alfv\'{e}n speed $\vA$ appears as mostly constant, with $\vA \sim 10~$km s$^{-1}$ for a large range of number densities from 0.1 to $100 \mathrm{cm}^{-3}$. 
This reflects the same scaling $B \propto \rho^{1/2}$, predicted by theory and simulations \citep{padoan_superalfvenic_1999} on small scales, deep inside molecular clouds. Note that there is however significant scatter that covers the classical adiabatic compression relation $B \propto \rho^{2/3}$ \citep{crutcher_magnetic_2010}.

Why is the Alfv\'{e}n speed roughly constant across scales? We propose here two mechanisms working in tandem. First, on very large scales, the magnetic energy saturates at a fixed fraction of the turbulent kinetic energy. This implies $B^2 \propto \rho \sigma_{\rm 1D}^2$, and is the classical argument of equipartition proposed by \citet{chandrasekhar_magnetic_1953, fiedler_ambipolar_1993, cho_anisotropy_2000, groves_radio-fir_2003}. Since the gas velocity dispersion is roughly constant throughout the disk (see Fig.~\ref{fig:velocity_compare}), this leads to both the expected scaling $B \propto \rho^{1/2}$ and the correct normalisation when using the average disk density $n_0 \simeq 1~\mathrm{cm}^{-3}$ and the average magnetic field $B_0 \simeq 10~\mu$G. 

What is less obvious is why this scaling relation persists inside dense clumps. The second, less classical explanation we propose here is the 2D magnetised collapse of these dense clumps from a razor thin (Toomre-unstable) disk. Assuming a tangential (or radial) magnetic field $B$ embedded in a collapsing thin cloud of constant height H and shrinking radius R, conservation of mass writes $M = \rho H \pi R^2$ and conservation of magnetic flux writes $\phi = B H 2R$. Combining these two relations while removing the radius, we obtain 
\begin{equation}
    B = \frac{\sqrt{\pi}}{2}\frac{\phi}{\sqrt{M H}} \sqrt{\rho} = B_0 \sqrt{n/n_0}
\end{equation}
assuming that the typical Toomre-unstable clump has initial radius $R\simeq H$, initial density $\rho=\rho_0$ and initial magnetic field $B=B_0$. This two-step model explains why the Alfv\'{e}n speed remains constant from the largest scales in the disk down to collapsing molecular clouds. Furthermore it remains constant at even smaller scales, as advocated by the work of \cite{padoan_superalfvenic_1999}, although these scales are for us unresolved and captured only by our sub-grid model.

\begin{figure*}
    \includegraphics[width=0.9\linewidth]{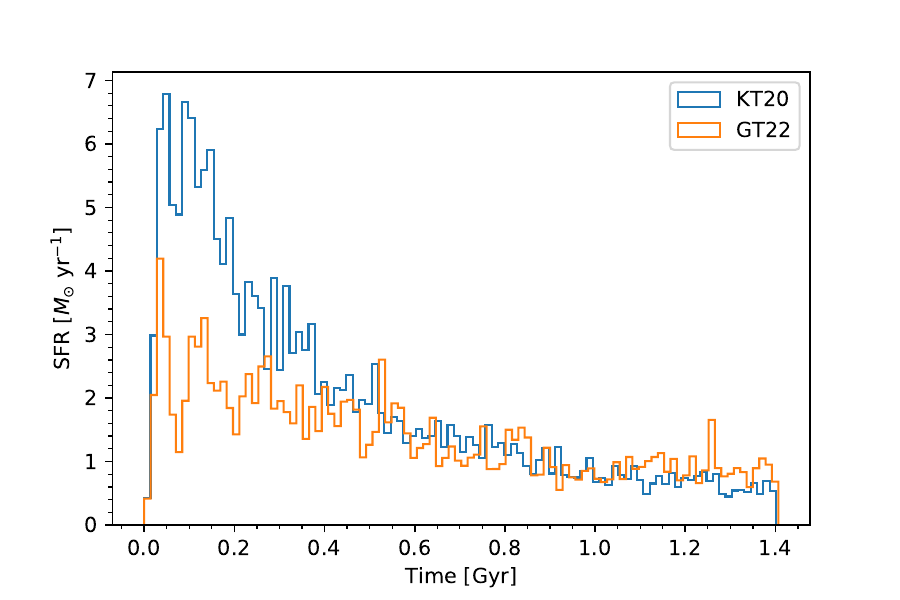}
    \caption{A comparison of the star formation history in both simulated galaxies, using either the non-magnetised sub-grid star formation recipe from \citet{kretschmer_forming_2020} (blue) or our magnetised sub-grid recipe (orange).}
    \label{fig:sfh_plots}
\end{figure*}
\begin{figure*}
\includegraphics[width=\linewidth]{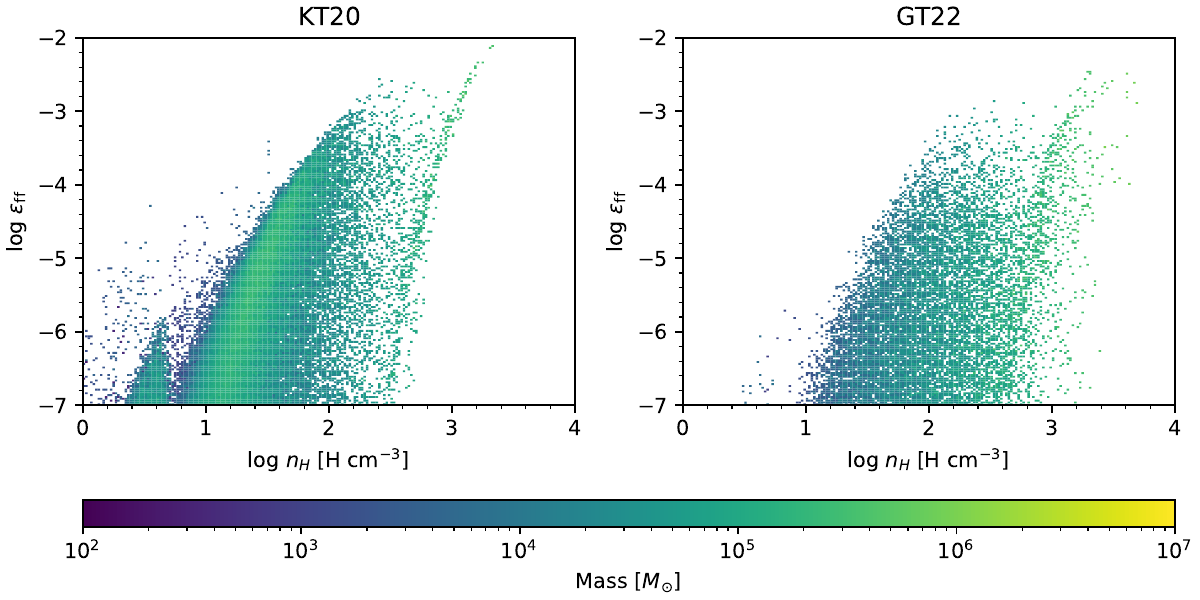}
\caption{Mass-weighted 2D histograms of efficiency per free-fall time $\epsilon_{\rm ff}$ versus number density $n_{\rm H}$, taken from a snapshot at 1.4 Gyr. The left panel is from our control simulation with the non-magnetised sub-grid star formation recipe, while the right panel is from our simulation that incorporates the magnetised model.}
\label{fig:eff_plots}
\end{figure*}

\subsection{Effect of our new star formation model}

Our sub-grid dynamo methodology allows the emergence of a strong magnetic field characterised by a roughly constant Alfv\'{e}n speed of magnitude $v_{\rm A} \simeq 10$~km/s. This translates into a plasma $\beta \simeq 10^{-3}$ inside dense star-forming molecular clouds in which the sound speed is roughly $\cs \simeq 0.2-0.4$~km/s. The nature of the turbulence inside the clouds is clearly dominated by the Alfv\'{e}nic Mach number $M_{\rm A} = \sigma_{\rm 1D}/v_{\rm A} \simeq 1-2$ as opposed to the sonic Mach number $M_{\rm s} = \sigma_{\rm 1D}/c_{\rm s} \simeq 10-20$, therefore we expect the magnetic field to be the dominant component in regulating star formation at small scales.  In particular, because the large-scale turbulent velocity dispersion and the Alfv\'{e}n speed are both roughly constant, as well as the sound speed inside molecular clouds, we expect an overall constant offset of the star formation efficiency.

The first difference between our magnetised star formation model and the non-magnetised model of \citet{kretschmer_forming_2020} can be seen in the star formation history of our isolated galaxy shown in Figure \ref{fig:sfh_plots}. In both cases, we observe an overall slow exponential decline with a prompt initial burst.
Note that this initial burst is considerably smaller than in other simulations of isolated galaxies \citep[see e.g.][]{stinson_star_2006} because we have included decaying turbulence in our initial conditions, preventing the formation of an unrealistic razor thin disk after the initial vertical collapse. 

Including the magnetic field reduces the initial SF rate from $5-6~\msun~\mathrm{yr}^{-1}$ to $2-3~\msun~\mathrm{yr}^{-1}$. As a result, the gas depletion time scale has been reduced by a factor of 2, resulting in a flatter SF history. At the end of the simulation, both models predict a SF rate around $1~\msun~\mathrm{yr}^{-1}$, but the magnetised case contains almost twice the amount of gas of the non-magnetised model by the end of the simulation.

In order to further understand the effect of the magnetic field on our star formation recipe, we have plotted in Figure~\ref{fig:eff_plots} the local star formation efficiency in each cell for both the non-magnetised case (left panel) and the magnetised case (right panel) as a function of the local gas density. Most notably our magnetised model results in the distribution of star-forming cells clearly shifting to higher densities, from $\sim 10~\rm cm^{-3}$ to $\gtrsim 100~\rm cm^{-3}$.  Overall however, the efficiency ranges themselves are quite similar in both cases, namely because the typical gas density is a factor of $\sim 2$ larger in the magnetised case.  This particular snapshot at 1.4 Gyr appears to have quite low efficiencies per free-fall time, with the non-magnetised and magnetised models reaching peak efficiencies of 0.7\% and 0.35\% respectively.  However, we emphasise this is only a snapshot and hence not representative of the larger star forming history; preceding snapshots demonstrate more active star formation, with the peak efficiency lying at 1\% for the densest cells in both model cases.

The total gas Schmidt law is represented in the left panel of Figure~\ref{fig:sigmaSFR_plots}, where we plot the gas surface density $\Sigma\gas$ against the SFR surface density $\Sigma\SFR$. $\Sigma\SFR$ is computed within a disk of radius 10~kpc and height 3~kpc, by azimuthally summing the mass of young stars (formed over the last 100 Myr) in radial bins of size 100~pc.  There is a slight difference in the power law index, where $n \sim 2.5$ and $n \sim 2$ for the non-magnetised and magnetised star formation models respectively.  The power law in the magnetised model however shows better agreement with observed values for individual spiral galaxies, which range from $n \sim 1.2 - 2.1$ \citep{wong_relationship_2002}.

The most striking difference in the left panel of Figure~\ref{fig:sigmaSFR_plots} is the  much higher gas surface density observed in the nuclear region. We reach $500~\msun {\rm pc}^{-2}$ using the magnetised star formation model against only  $100~\msun {\rm pc}^{-2}$. The particularly strong magnetic field in the nuclear region reduces the local star formation efficiency. This in turn requires more and denser gas to accumulate, in order to produce enough supernovae feedback to regulate the global star formation efficiency. This self-regulation mediated by feedback explains why our overall SF history is not strongly affected by our new model, while locally it is strongly modified. Interestingly, the magnetised star formation model allows also smaller values of the SFR at low surface densities. We see in the left panel of Figure~\ref{fig:sigmaSFR_plots} a tail of low SFR for surface densities below $10~\msun {\rm pc}^{-2}$. This tail is reduced for the non-magnetised star formation model, suggesting that the magnetic field in the outer regions is high enough with $B \simeq 3-4~\mu{\rm G}$ to also quench star formation there.

\begin{figure*}
\includegraphics[width=\linewidth]{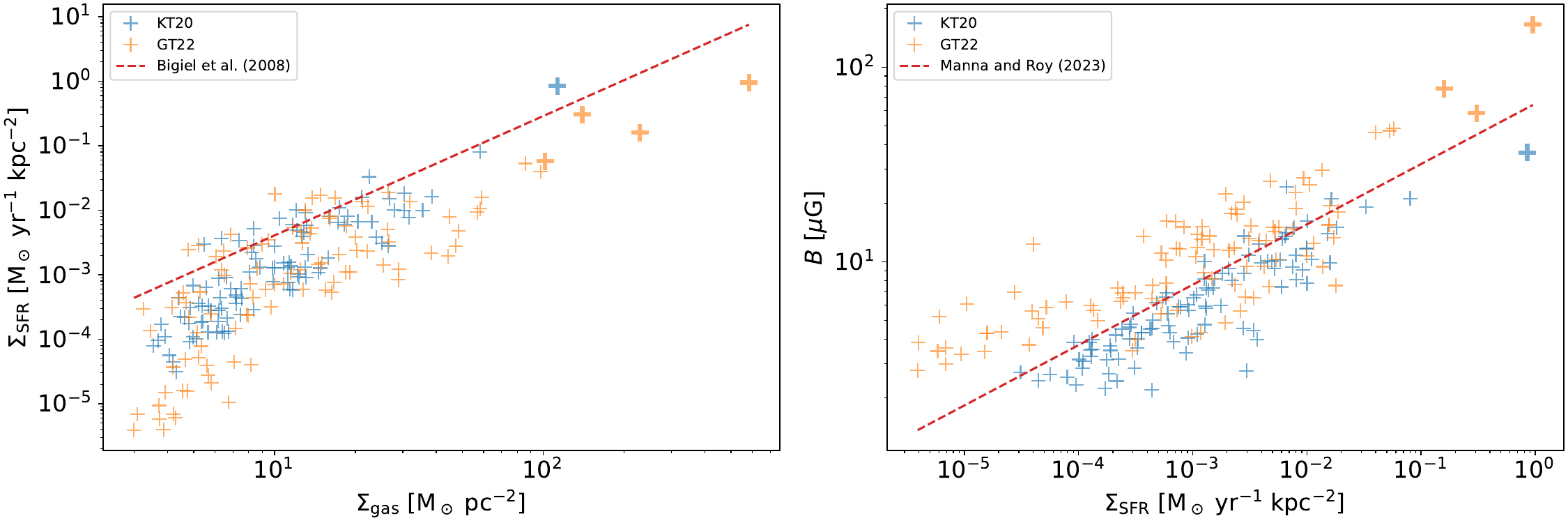}
\caption{Left: Gas surface density $\Sigma\gas$ and star formation rate surface density $\Sigma\SFR$ relation, within simulated galaxies using either the non-magnetised (blue) or magnetised (orange) sub-grid star formation models. The high gas surface density tail, which we attribute to the nuclear regions, is emphasised by thicker markers. Overplotted in the red dotted line is the average $\Sigma\gas-\Sigma\SFR$ law empirically found by \citet{bigiel_star_2008}.
Right: Star formation rate surface density $\Sigma\SFR$ and magnetic field strength $B$ relation, within simulated galaxies using either the non-magnetised (blue) or magnetised (orange) sub-grid star formation models.  Similarly, the high SFR surface density tail is additionally emphasised by thicker markers. Overplotted in the red dotted line is the $\Sigma\gas-B$ law empirically found by \citet{manna_magnetic_2023}.
In both subplots, we consider mass-weighted quantities within a disk around the centre of radius 10 kpc and height 3.2 kpc, split into 100 cylindrical bins. $\Sigma\SFR$ is computed assuming a stellar age of 100 Myr.}
\label{fig:sigmaSFR_plots}
\end{figure*}

\section{Discussion and comparison to observations}\label{sec:discussion}

To assess the consistency of our simulation results with observations, we compare the star formation rate surface density $\Sigma\SFR$, the total gas surface density $\Sigma\gas$, and the volume-weighted average magnetic field strength $B$ to observed relations found in literature.  The left panel of Figure \ref{fig:sigmaSFR_plots} displays the average total gas Schmidt law as fitted by \citet{bigiel_star_2008}, for a sample of 18 nearby galaxies at sub-kpc resolution.  Similar to our analysis, they provide values for $\Sigma\SFR$ and $\Sigma\gas$ across azimuthally-averaged radial profiles.  While our non-magnetised and magnetised star formation models appear to under-predict their averaged power law, they both lie well within the observed spread around this power law.  \citet{bigiel_star_2008} emphasise that $\Sigma\SFR$ and $\Sigma\gas$ differ largely across the galaxies in their sample, with the average  $\Sigma\SFR$ possessing an RMS scatter of $\sim 0.3$ dex.  In fact, where some galaxies are well described by a single power law, others are not.  Hence we wish to make the key point here that our simulations necessarily align with the generally observed range of values in SFR-gas space.

Furthermore, we note that the total gas Schmidt relations of both simulation models appear to possess a "knee" around the value of $\Sigma\gas \sim 10~M_\odot~\rm pc^{-2}$.  Both \citet{bigiel_star_2008} and previous studies \citep{morris_study_1978, martin_star_2001, wong_relationship_2002} have demonstrated how the HI radial profiles in spirals reflect a remarkably consistent saturation at high column densities, with a threshold of $\Sigma\gas \approx 9~M_\odot~\rm pc^{-2}$.  H$_2$ surface density, on the other hand, generally reflect no similar limitations.  Therefore the apparent difference between the $\Sigma\gas-\Sigma\SFR$ relation at lower and higher column densities, as opposed to a single power law, arguably reflects a transition between an HI-dominated ISM to an H$_2$-dominated ISM that is consistent with observations.

We additionally compare the relation between magnetic field strength and $\Sigma\SFR$ produced by both models, to that inferred from radio continuum observations and reported in the recent work of \citet{manna_magnetic_2023}. In the right panel of Figure \ref{fig:sigmaSFR_plots}, we show their average relation found using a sample of 11 nearby galaxies is in striking agreement with our simulations.  Interestingly, our non-magnetised star formation model appears to result in central magnetic fields that are too weak. Previous semi-analytical models of the $B-\Sigma\SFR$ relation, such as that proposed by \citet{schleicher_new_2013}, similarly predict central magnetic field values that are too weak by a factor of 2 to 3.  As \citet{manna_magnetic_2023} discuss, a key parameter of their model is the fraction of turbulent kinetic energy that is converted to magnetic energy, otherwise known as the saturation level of the turbulent dynamo ($f_{\rm sat}$).  While \citet{schleicher_new_2013} assume an $f_{\rm sat} \sim 5$\%, in our case (see Figure \ref{fig:velocity_compare}) we find a significantly larger value of $f_{\rm sat} \simeq 25$\%. This could partly explain why our simulations are in better agreement with observations. 

\section{Conclusions}

In this paper, we have performed ideal MHD simulations of an isolated, Milky
Way-like galaxy using a new sub-grid model for our star formation recipe that accounts for the presence of magnetic fields. For this, we used a Large Eddy Simulation approach to compute the velocity dispersion of the sub-grid turbulence, with the same LES model used to power a sub-grid turbulent dynamo. This mean-field $\alpha$ dynamo allows us to amplify our vanishingly small initial magnetic field to a value close to (but smaller than) equipartition. Our star formation model is based on the Sub-grid Scale (SGS) multi-freefall model proposed originally by \cite{federrath_star_2012}, using both the sub-grid turbulence and the magnetic field computed self-consistently by the \texttt{RAMSES} code, and extends numerical methods described in \citet{kretschmer_forming_2020}. As discussed in Section~\ref{sec:model}, the effect of the magnetic field is to both reduce the width of the log-normal gas density PDF and make the collapse criterion for molecular cores more difficult to achieve. Overall, as shown in this paper, the efficiency of star formation at small, sub-grid scales, can be significantly reduced if the magnetic field strength is high enough. 

Interestingly, we find that the Alfv\'en speed is remarkably constant from galactic scales down to molecular cloud scales. It value is of the order of 10~km/s, corresponding exactly to the average sound speed of the warm gas, but smaller that the average velocity dispersion, which in our case is closer to 20~km/s. 
The origin of this remarkable uniformity of the Alfv\'en speed is due to 2 combined effects: 1- the saturation of the turbulent dynamo at a constant ratio of magnetic to turbulent energy, 2- the 2D collapse of razor thin molecular cloud with a magnetic field aligned with the disk. 

Including the magnetic field in the multi-free fall star formation model reduces the overall initial  star formation rate by nearly a factor of 2 when compared to the non-magnetised reference case.  However, star formation somehow self-regulates so that the final SFR is around $1~\msun~\mathrm{yr}^{-1}$ after $\gtrsim 1$ Gyr in both cases. The most striking effect of reducing the local SF efficiency at small scales is to compress the gas to a much higher density (and field strength) by a factor of up to 5 in the nuclear region. As a result, the local SF efficiency does not change much compared to the non-magnetised case, and lies close to the value adopted in most galaxy formation simulations, $\eff \sim 0.01$ \citep[see][]{krumholz_slow_2007,mckee_theory_2007}. The dispersion of the local efficiency is also in good agreement with observations of nearby galaxies, where $\eff = 0.003 - 0.026$ with a median value around 0.007 \citep{utomo_star_2018}. 

Our results compare well with the mean SFR surface density - magnetic field relation observed in nearby galaxies \citep{manna_magnetic_2023}. Although encouraging, our results are still lacking many realistic aspects of galaxy formation and cosmic magnetism. Because our initial magnetic field was vanishingly small, only the turbulent dynamo is at work here. As a result, our mean toroidal field is very small and the magnetic field geometry is dominated by the fluctuating component. We plan to explore in the future different initial field configurations, including ones with a strong pre-existing toroidal field. This could have an additional impact on the overall star formation efficiency.

Another important caveat of this model is that we consider the effect of the magnetic field only through the magnetic pressure, completely neglecting anisotropic effects. However as shown in the recent work of \cite{beattie_multi-shock_2021}, the gas density distribution in sub-Alfv\`enic turbulence is highly anisotropic. Their theoretical model could be used to refine our model and account for this anisotropy.

Furthermore, as this study focuses on isolated galaxy simulations, we cannot account for the effect of the cosmological environment (inflows) on the galactic magnetic field topology and star formation.  Past studies agree that on average, isolated galaxies have higher contents of HI and lower star formation rates than galaxies residing within clusters \citep[see][and references therein]{boselli_environmental_2006}. Alternatively, motions in a forming galaxy cluster combined with turbulence in the intracluster gas is believed to amplify the magnetic field in cluster galaxies still undergoing major mergers \citep{roettiger_magnetic_1999,roettiger_cluster_1999, subramanian_evolving_2006}. In future work, we intend to examine our star formation model with these factors in mind, starting with more comprehensive simulations of isolated galaxies that can be followed up by those of larger-scale galaxy clusters.

\section*{Acknowledgements}
We thank the anonymous reviewer for their insightful comments, which helped us improve the quality of this paper.  The simulations included in this work were executed on computational resources managed and supported by Princeton Research Computing, a consortium of groups including the Princeton Institute for Computational Science and Engineering (PICSciE) and the Office of Information Technology's High Performance Computing Center and Visualization Laboratory at Princeton University. E.G. acknowledges support from a Predoctoral Fellowship administered by the National Academies of Sciences, Engineering, and Medicine on behalf of the Ford Foundation, and the National Science Foundation Graduate Research Fellowship Program under grant number DGE-2039656. Any opinions, findings, and conclusions or recommendations expressed in this material are those of the authors and do not necessarily reflect the views of these funding organisations.

\section*{Data Availability}
The data underlying this article will be shared on reasonable request to the corresponding author.



\bibliographystyle{mnras}
\bibliography{bibliography} 




\appendix




\bsp	
\label{lastpage}
\end{document}